\documentclass[aps,prl,twocolumn,superscriptaddress,%showpacs %twocolumn
,floatfix]{revtex4}
\usepackage{layouts}

\usepackage{amsmath}
\usepackage{amsfonts}
\usepackage{amssymb}
\usepackage{graphicx}
\usepackage{color}
\usepackage[colorlinks=true,citecolor=blue,linkcolor=blue,urlcolor=blue]{hyperref}
\usepackage{times}
\usepackage{amstext}
\usepackage{latexsym}
\usepackage[running,mathlines]{lineno}
\usepackage{verbatim}
\usepackage{physics}

\providecommand{\hc}[1]{#1^{\dagger}}
\providecommand{\HC}{\mathrm{h.c.}}
\providecommand{\norm}[1]{\lVert #1 \rVert}

\providecommand{\ket}[1]{\lvert #1 \rangle}

\newcommand{\im}{\mathrm{i}}
\newcommand{\E}{\mathrm{e}}

\begin{document}
\raggedbottom

\title{Impact of temporal correlations, coherence, and postselection on two-photon interference}

\author{Fernando Redivo Cardoso}
\affiliation{Departamento de F\'{i}sica, Universidade Federal de S\~{a}o Carlos, 13565-905 S\~{a}o Carlos, S\~{a}o Paulo, Brazil}
\affiliation{Department of Physics, Stockholm University, 10691 Stockholm, Sweden}

\author{Jaewon Lee}
\affiliation{Department of Physics, Stockholm University, 10691 Stockholm, Sweden}

\author{Riccardo Checchinato}
\affiliation{Department of Physics, Stockholm University, 10691 Stockholm, Sweden}

\author{Jan-Heinrich Littmann}
\affiliation{Department of Physics, Stockholm University, 10691 Stockholm, Sweden}

\author{Marco De Gregorio}
\affiliation{Technische Physik, Physikalisches Institut and W\"urzburg-Dresden Cluster of Excellence ct.qmat, Universit\"at W\"urzburg, Am Hubland, D-97074 W\"urzburg, Germany}

\author{Sven H\"{o}fling}
\affiliation{Technische Physik, Physikalisches Institut and W\"urzburg-Dresden Cluster of Excellence ct.qmat, Universit\"at W\"urzburg, Am Hubland, D-97074 W\"urzburg, Germany}

\author{Christian Schneider}
\affiliation{Institut of Physics, University of Oldenburg, D-26129 Oldenburg, Germany}

\author{Celso J. Villas-Boas}
\affiliation{Departamento de F\'{i}sica, Universidade Federal de S\~{a}o Carlos, 13565-905 S\~{a}o Carlos, S\~{a}o Paulo, Brazil}

\author{Ana Predojevi\'{c}}
\email{ana.predojevic@fysik.su.se}
\affiliation{Department of Physics, Stockholm University, 10691 Stockholm, Sweden}

%\email{}

\begin{abstract}
Two-photon interference is an indispensable resource in quantum photonics, but it is not straightforward to achieve. The cascaded generation of photon pairs contains intrinsic temporal correlations that negatively affect the ability of such sources to perform two-photon interference, thus hindering applications. We report on how such correlation interplays with decoherence and temporal postselection, and under which conditions temporal postselection could improve two-photon interference visibility. Our study identifies crucial parameters and points the way to a source with optimal performance.
\end{abstract}

\maketitle

Photons are a highly suitable choice for flying qubits due to their ease of generation, ability to carry encoding in various degrees of freedom, and low decoherence resulting from low interaction with the environment. The latter comes at a cost of a limited means to enable photons to interact. The primary method for achieving photon interaction is through two-photon interference on a beamsplitter, also known as the Hong-Ou-Mandel interference \cite{Hong1987}. The use of this effect spans numerous platforms and applications, including operations such as teleportation \cite{Bouwmeester} and entanglement swapping \cite{Zukowski}, linking quantum systems \cite{Beugnon, Krutyanskiy}, photonic circuits \cite{Wang2020}, fusing of photonic states \cite{Cogan}, and state control and characterization \cite{Walker}. 

Spontaneous parametric downconversion and quantum dots are the two most commonly used systems to generate bipartite  photon entanglement. In the case of spontaneous parametric downconversion, the high two-photon interference contrast is engineered by modifying the joined-spectral amplitude \cite{Grice} and eliminating the underlying correlations by means of spectral filtering. In contrast to %photon-pair generating process such as 
spontaneous parametric downconversion, where the two photons are generated simultaneously, a quantum dot emits a pair of photons as a time-ordered cascade, with the biexciton photon preceding the exciton photon. The resulting two-photon wave function describing the cascade emission has the following form \cite{Huang1993, simon2005} 

\begin{equation}\label{single cascade wavefunction}
    \psi(t_b,t_x) = 2 \sqrt{\Gamma_b \Gamma_x} \E^{-\Gamma_b t_b} \Theta(t_b) \E^{-\Gamma_x (t_x - t_b)} \Theta(t_x - t_b),
\end{equation}
where $t_{b}$ ($t_{x}$) is the emission time of the biexciton (exciton) photon, while $\Gamma_{b}$ ($\Gamma_{x}$) denotes the biexciton (exciton) decay rate. The factor $\Theta(t_x - t_b)$ has the form of a Heaviside step function and accounts for the temporal ordering (cascade emission) that induces correlations between the emitted photons \cite{Huang1993, simon2005}. The extent of the correlations between the two photons, and hence, the purity of a single photon belonging to a pair, can be quantified by determining the trace of the squared reduced density operator \cite{Zyczkowski}. Due to the form of the two-photon wave function,  $\psi(t_b,t_x)$, tracing over the  biexciton (exciton) photon subsystem will reveal that the exciton (biexciton) is not in a pure state. For example, by denoting $\rho_x = \Tr_{b}(\psi^*\psi)$, we obtain \cite{simon2005} the purity of the exciton photon as

\begin{equation}\label{trace reduced rho}
    \Tr(\rho_x^2) = \frac{\Gamma_b}{\Gamma_b + \Gamma_x}.
\end{equation}
The upper bound of $\Tr(\rho_x^2)$ is unity and manifests itself in the case the state is pure. Hence, $\Gamma_x \ll \Gamma_b$ implies high purity, a condition required to achieve high visibility contrast in two-photon interference experiments. On the other hand, $\Gamma_x \gg \Gamma_b$ indicates that the temporal correlations described by $\psi(t_b,t_x)$ are strong and that they would lead to a reduction of purity of the individual photons pertaining to a photon pair.

In a quantum dot embedded in a bulk material, the biexciton and exciton decay times are comparable \cite{Sek2010, Bacher1999}.
 As a result, the two-photon interference visibility observed in the experiments rarely exceeded 0.5 \cite{Rota2020}. On the other hand, the ability to use photon pairs generated by a single quantum dot in experiments that rely on two-photon interference \cite{Bouwmeester, Zukowski} is an essential prerequisite for their use. Therefore, there is a need for an in-depth analysis of the origins of the low interference contrast and mitigation strategies. Here, we theoretically and experimentally investigate how the visibility of the two-photon interference is affected by temporal correlations, decoherence, and the temporal postselection. Our results identify the strategy required to maximize the visibility.

The measurements were performed using an In(Ga)As quantum dot embedded in a micropillar cavity. The cavity was designed to feature a low quality factor (200-300) and in return provide a bandwidth of $\sim$5\,nm \cite{laia2022}. The quantum dot was excited resonantly by means of two-photon resonant excitation of the biexciton \cite{jayakumar2013}. To this end, we employed an excitation laser featuring 80\,MHz repetition rate and a pulse length of 15\,ps. The excess laser scattering was removed by means of spectral and polarization filtering. The biexciton and exciton single photons were separated using a diffraction grating and coupled into single mode fibers. The low multi-photon contribution in the quantum dot emission was confirmed by measuring the auto-correlation function (shown in \cite{sup}). The measurements yield $g^{(2)}_{b}(0)$=0.0144(19) for biexciton and $g^{(2)}_x(0)$=0.0074(11) for exciton photons. 
%To estimate the upper bound on purity of the individual photons 
We also performed lifetime measurements and obtained $\tau_b=237.16(59)$\,ps and $\tau_x=367.61(99)$\,ps for the biexciton and exciton, respectively  (data plots and the respective fits are given in \cite{sup}). The ratio of the lifetimes indicates that the correlations between the biexciton and the exciton emission time should significantly reduce the purity of the individual photons \cite{Huber13}. However, to determine the upper bound of the two-photon interference visibility we had to employ a theoretical model.

\begin{figure}[t]
\includegraphics[width=0.95\linewidth]{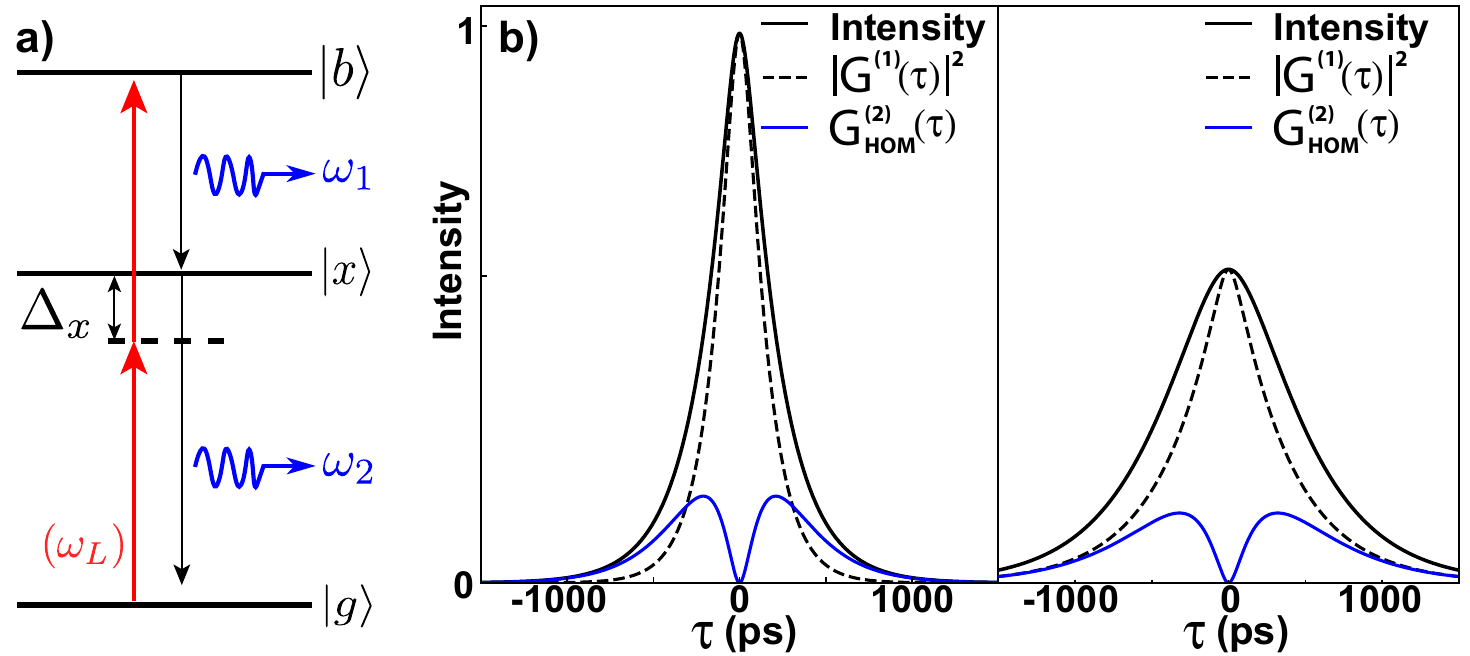} 
\caption{(a) Schematic of the three-level system. A laser with a frequency $\omega_L$ coherently couples the ground state, $\ket{g}$, and the biexciton, $\ket{b}$, via a virtual level. Upon excitation, the biexciton photon, $\omega_1$, and exciton photon, $\omega_2$, are emitted. The two-photon resonance is detuned from the single-photon resonance (ground state, $\ket{g}$, to exciton, $\ket{x}$) by $\Delta_x$. (b) The results of the sensor method simulation: intensity, first-order correlation function, $G^{(1)}(\tau)$ and second-order correlation function, $G^{(2)}_{HOM} (\tau)$, for the biexciton (left) and the exciton (right). For optimal comparison of the two plots, the area under the intensity peak was scaled to unity. The simulation assumed the experimentally determined biexciton and exciton lifetimes ($\tau_b=237.16(59)$\,ps and $\tau_x=367.61(99)$\,ps) and no dephasing. These results yield two-photon coincidence probability of $P_{0}=0.15$ while the visibility is equal to $\nu$=0.54  for both the biexciton and the exciton photon.} 
\label{Fig1}
\end{figure}

We modeled the quantum dot as a three-level system consisting of a ground state, exciton, and biexciton, as shown in Fig. \ref{Fig1}a. Once the quantum dot is two-photon resonantly excited it decays to the ground state via emission of photons with frequencies $\omega_1$ (biexciton) and $\omega_2$ (exciton). The full description of the system dynamics is given in \cite{sup}. 

%%%%%%%%%%%%%.   FIRST  FIGURE !   %%%%%%%%%%%%%%%%%

Two indistinguishable photons impinging on the beamsplitter will undergo interference \cite{hong1987}. If we denote the beamsplitter input modes as $a$ and $b$, the time-resolved interference of two photons emitted by independent sources is given by \cite{Legero, Kiraz2004}

\begin{multline}\label{g2 two-time 2}
%\begin{split}
    G^{(2)}_{HOM} (t,\tau) = \frac{1}{2} \left[ \expval{\hc{\xi}_a(t) \xi_a(t)} \expval{\hc{\xi}_b(t+\tau)\xi_b(t+\tau)} \right.  \\ \left. - 2 \Re{\left(G^{(1)}_a(t,\tau)\right)^* G^{(1)}_b(t,\tau)} \right].
%\end{split}
\end{multline}
In the equation above the $n_{a}(t)=\expval{\hc{\xi}_{a}(t) \xi_{a}(t)}$  and $n_{b}(t+\tau)=\expval{\hc{\xi}_{b}(t+\tau) \xi_{b}(t+\tau)}$ are values of photon number in modes a and b and, as such, are proportional to the intensity in the respective mode. On the other hand, $G^{(1)}_{a,b}(t,\tau)$ is the first-order (field) correlation function. The function describing the two-photon interference is the second-order (intensity) correlation function $G^{(2)}_{HOM}(t,\tau)$. 

%%%%%%%%%%%%%.   FIRST  FIGURE !   %%%%%%%%%%%%%%%%%

To calculate the first-order correlation function, we implemented the sensor method \cite{delvalle2012}. This approach is based on supplementing the three-level system by weakly coupled quantized radiation modes (two-level systems) that act as sensors. Here, we introduced two such sensors (one per emission frequency) and they are described by the following Hamiltonian

\begin{equation}\label{Hamiltonian_sensors}
    H_s = \sum_{j=1}^{2} \left\{ \omega_j \hc{\xi}_j \xi_j + g \left[ \left( \sigma_{xb} + \sigma_{gx} \right) \hc{\xi}_j + \HC \right] \right\}, 
\end{equation}
where $\omega_j$ is the resonance frequency, while $\hc{\xi}_j$ $(\xi_j)$ correspond to the creation (annihilation) operator for a sensor $j$. The coupling strength between the sensors and the quantum dot is given by $g$. The presence of the sensors must not perturb the three-level system by, for example, introducing back action from the sensed excitation. Therefore, the parameter $g$ must be very small (we consider $g=10^{-3}$). The detailed description of the sensor method is given in \cite{sup}.

The sensor method allows the treatment of photon correlations taking into account the uncertainties in time and frequency of the detection \cite{delvalle2012}.  Hence, it is the ideal approach to quantify the correlations present in quantum dot emission. We employed it to calculate the photon number and the first-order correlation function of the sensor modes. By replacing these in \eqref{g2 two-time 2} and integrating over $t$, we obtain the $G^{(2)}_{HOM} (\tau)$. Similarly, by integrating the individual terms in \eqref{g2 two-time 2}  one obtains the correlation in photon number and $G^{(1)} (\tau)$, respectively. The results are shown in Fig. \ref{Fig1}b. They yield the probability of a coincidence at the outputs of the beamsplitter of $P_{0}=0.15$. While this value is significantly lower than the classical limit of $P_{\infty}=0.5$, imposed by full absence of the interference effect, it demonstrates the detrimental effect of the temporal correlations. The corresponding visibility $\nu$ is 0.54, where the visibility is defined as $(P_{\infty}-P_{0})/(P_{\infty}+P_{0})$ \cite{Kaltenbaek2006}.

%%%%%%%%%%%%%.    FIGURE 2   %%%%%%%%%%%%%%%%%

\begin{figure}
\centering
\includegraphics[width=0.94\linewidth]{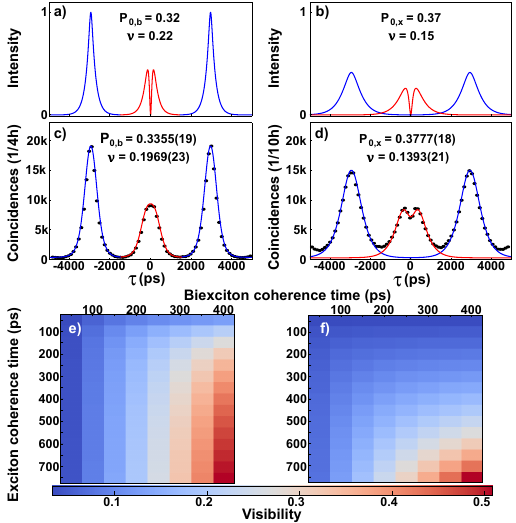}
\caption{The results of the sensor method calculation for a) biexciton and b) exciton. We assumed the coherence time of biexciton and exciton to be $\tau_{cb}$=200\,ps and $\tau_{cx}$=450\,ps, respectively. The $G^{(2)}_{HOM} (\tau)$ is plotted as the red central peak, while in blue we plotted the profile of the peak that would be achieved if no interference occurs. $P_{0}$ and $\nu$ match the experimentally observed ones. Panels c) and d) show the results of the two-photon interference measurement for biexciton and exciton, respectively. The measurement data are shown as points. The solid line is theory prediction achieved by convolution of the curves shown in the panel a) and b) with the response of the detector. The panels e) and f) show the visibility of the biexciton and exciton photon, respectively, for various values of the biexciton and exciton photon coherence time. The numerical values are given as tables in \cite{sup}.}
\label{fig:Fig2}
\end{figure}

%%%%%%%%%%%%%.    FIGURE 2   %%%%%%%%%%%%%%%%%

We accessed $P_{0}$ and $\nu$ experimentally. To achieve this we employed two unbalanced interferometers with a nominal delay of 3\,ns. The first interferometer served to generate two laser pulses required to excite the quantum dot, while  we used the second interferometer to observe two-photon interference of photons emitted in two consecutive excitations. The detailed schematic of the setup is given in \cite{Huber15}. The results of these measurements are shown as points in Fig. \ref{fig:Fig2}c and Fig. \ref{fig:Fig2}d for biexciton and exciton, respectively. The measurements yield $P_{0,b}=0.3355(19)$ and $P_{0,x}=0.3777(18)$ for biexciton and exciton, respectively. The values $P_{0,b}$ and $P_{0,x}$ were determined by summing the area under the central peak and dividing it with the sum of the areas of the adjacent peaks \cite{Santori2002}. In the absence of the two-photon interference the three peaks would be identical, resulting in $P_{\infty}=0.5$. The corresponding values of the visibility were found to be $\nu_{b}=0.1969(23)$ and $\nu_{x}=0.1393(21)$, respectively. These values suggest that the  reduction of the interference contrast is not solely caused by the cascaded emission, but that there is a coexisting effect of the dephasing of the quantum dot levels. However, accessing the degree of dephasing for a cascaded decay is not straightforward. Specifically, when a quantum dot is driven in the two-level regime, exciting only the exciton, the information on the dephasing is commonly extracted from the two-photon interference measurement \cite{Ding16, Unsleber}. This approach is motivated by slow dephasing mechanisms that cause methods such as the $G^{(1)}$ measurement using Michelson interferometer to record a higher degree of dephasing than commonly observed in a measurement implemented using photons emitted shortly after each other \cite{Thoma}. In a three-level system, two-photon interference is reduced by both the dephasing of any of the energy levels involved and cascade correlations. Therefore, estimating dephasing-induced photon distinguishability is more complex than in the two-level regime. 

However, accessing the coherence time and purity of the photons pertaining to a cascade is crucial for experiments relying on two-photon interference. Therefore, we investigated this interplay between cascade-correlations and decoherence-induced reduction of two-photon interference contrast. The sensor method can also be used for this purpose. We introduced the dephasing via the Lindblad terms of the master equation \cite{sup}. This enabled us to determine $G^{(2)}_{HOM} (\tau)$ for any value of the coherence time attributed to either biexciton or exciton. Several examples of $G^{(2)}_{HOM}(\tau)$ we calculated using this method are plotted in Fig. 3 in \cite{sup}, while the results that most closely match the experimentally observed values of $P_{0,b}$ and $P_{0,x}$ are shown in Fig. \ref{fig:Fig2}a and \ref{fig:Fig2}b, respectively. To fit the $G^{(2)}_{HOM}(\tau)$ calculated using sensor method shown in Fig. \ref{fig:Fig2}a and \ref{fig:Fig2}b to the experimental data, the $G^{(2)}_{HOM}(\tau)$ needs to be convoluted with the response of the detector employed in the measurement, resulting in curves shown in Fig. \ref{fig:Fig2}c and Fig. \ref{fig:Fig2}d.  The data fitting confirmed the values of coherence time of $\tau_{cb}$=200(25)\,ps and $\tau_{cx}$=450(25)\,ps for biexciton and exciton, respectively. We used this result to determine the dephasing times of the biexciton and exciton, which are 346(74)\,ps and  1160(167)\,ps, respectively. In addition, the coherence time of the biexciton and exciton are, as anticipated, significantly longer than the $G^{(1)}$ values we measured employing the Michelson interferometer, which yielded 193(9)\,ps and 130(5)\,ps for biexciton and exciton, respectively.

We determined the values of the expected two-photon interference visibility for a wide range of different values of the coherence time. The results are shown in Fig.\ref{fig:Fig2}e and Fig.\ref{fig:Fig2}f for biexciton and exciton, respectively. They show that the visibility for the biexciton photon is unequally affected by the coherence loss of the biexciton and the exciton. The numerical values of the visibility are given as tables in \cite{sup}. Furthermore, to achieve a generalized analysis of the problem, we determined the values of the two-photon interference visibility for the same wide range of coherence time values and a biexciton to exciton lifetime ratio of 1:2 (200\,ps and 400\,ps). The plots and the table with numerical values are given in \cite{sup}.

The most common method for eliminating undesired correlations between a pair of entangled photons is postselection. While sources based on spontaneous parametric downconversion rely on spectral postselection \cite{Grice}, the system we are addressing here asks for temporal postselection. In this scenario, to improve the purity of individual photons biexciton decay is truncated. We conducted both theoretical and experimental study of this scenario. 

To theoretically address the effect of temporal postselection on two-photon interference visibility, we employed the quantum trajectory approach \cite{Carmichael}. This method determines single quantum trajectories from the time evolution of the non-hermitian Hamiltonian,

\begin{equation}\label{eq: H_nh}
    H' = H_I - \frac{\im \hbar}{2} \sum_k \hc{C}_k C_k,
\end{equation}
where $C_k$ corresponds to the $k$-th collapse operator and $H_I$ is the time-independent Hamiltonian describing our three level system together with the sensors (eq. (16) \cite{sup}). Each trajectory consists of a continuous evolution governed by $H'$ and a quantum jump that takes place at a random time, enabling spontaneous emission. The continuous evolution is described by the operator $U(t) = e^{-iH't/\hbar}$, where the imaginary term of $H'$ decreases the norm of the state vector such that $\norm{\ket{\psi(t_1)}} > \norm{\ket{\psi(t_2 > t_1)}}$. To describe the quantum jumps we discretized the time evolution and for each step we computed the norm of the state vector. We compared values of the norm of the state vector with a randomly generated number $r$ ($0\leq r \leq 1$). We assumed that when the condition $\norm{\ket{\psi(t_1)}} < r$ was satisfied the collapse and the re-normalization of the wavefunction took place. 

%%%%%%%%%%%%%% Figure 3 !!!!!!%%%%%%%%%%%%%%%%%%%%%%%%%%%%%

\begin{figure}[t]
    \centering
     \includegraphics[width=0.95\linewidth]{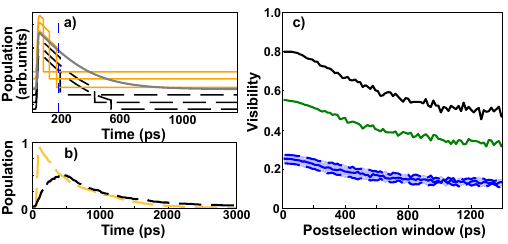}
    \caption{a) The trajectories shown in orange exhibit a quantum jump before the 180\,(ps) cut-off (blue dashed line), while the trajectories shown in black feature a quantum jump after the cut-off time. In gray is plotted the change of biexciton population with time calculated as the sum of all possible trajectories. The intensity of the individual plots is given in arbitrary units as the individual curves were vertically displaced for better visual presentation. b) The population change with time for exciton when the corresponding biexiton had a postselection cut-off at at 512\,ps (black) and 64\,ps (orange). c) Theoretically predicted result of postselection for no-dephasing (black), dephasing such that the coherence time of biexciton and exciton is at 80$\%$ of the Fourier-transform limit (green). The parameters of lifetime were equal to the ones of the emitter we used in the experiment. In blue is shown the result for our emitter, accounting for the coherence time determined experimentally. The light blue region accounts for the error with which we have determined the coherence time.} 
    \label{fig:Fig3}
\end{figure}

%%%%%%%%%%%%%% Figure 3 !!!!!!%%%%%%%%%%%%%%%%%%%%%%%%%%%%%

While a single trajectory corresponds to the single evolution of the initial state under a random collapse condition, an averaged set of $N$ trajectories ($N\rightarrow\infty $) approximates the density operator $\rho(t)$ typically obtained by solving the ensemble quantum master equation. Thus, with a numerical solution for a large number of trajectories, we can simulate the effect of temporal postselection based on the biexciton emission that occurs before a given time (Fig. \ref{fig:Fig3}a). This postselection procedure modifies the shape of the corresponding exciton wavepacket (Fig. \ref{fig:Fig3}b). The postselection effectively truncates the biexciton wavepacket, increasing the purity of the exciton-reduced density matrix. The increase in visibility as a result of the temporal postselection is shown in the Fig. \ref{fig:Fig3}c.

\begin{figure}[t]
    \centering
     \includegraphics[width=0.95\linewidth]{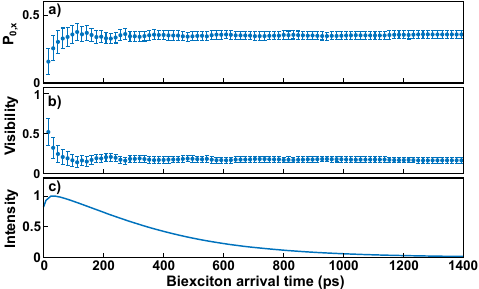}
    \caption{a) The probability of a coincidence at the output of the beamsplitter for exciton photon $P_{0,x}$ and b) the visibility, respectively. Both of these results were achieved in a measurement conditioned by biexciton detection and plotted as a function of the size of the postselection window. c) Lifetime of the biexciton used to define the postselection window.}
    \label{fig:Fig4}
\end{figure}

To test the effect of postselection experimentally, we conditioned the detection of exciton two-photon interference on the detection of a biexciton photon. This allowed us to use the time of biexciton photon detection as the postselection criterion. The results are shown in Fig. \ref{fig:Fig4}. The observed effect of postselection is stronger than predicted theoretically, especially for short postselection windows. However, it should be noted that the accuracy of the result is limited by the measurement statistics for such a short postselection window.

%%%%%%%%%%%%%% Figure 4 !!!!!!%%%%%%%%%%%%%%%%%%%%%%%%%%%%%

%%%%%%%%%%%%%% Figure 4 !!!!!!%%%%%%%%%%%%%%%%%%%%%%%%%%%%%

The concept of correlation of photons emitted as a cascade can be generalized to the phenomenon of time-energy entanglement as introduced by Franson \cite{Franson1989}. The manifestation of Franson's interference is based on two properties of the emission time: the uncertainty of when a cascade is emitted and the strong correlations of the photons belonging to a cascade. This property of quantum dot emission has recently been exploited to demonstrate time-energy entanglement \cite{Hohn2023}. However, any other application of entangled photon pairs generated by a quantum dot requires the complete elimination of the cascade-induced correlations.

We have analysed the combined effect of decoherence and cascade-induced correlations on the visibility of the two-photon interference of biexciton (exciton) photons emitted sequentially from a semiconductor quantum dot. We have shown that the sensor method and the two-photon interference measurement provide access to the coherence times of the biexciton and exciton photons. We generalised our results and showed that, in terms of two-photon interference visibility, the biexciton and exciton photons do not respond equally to the loss of coherence. We also investigated temporal postselection as a method to improve the visibility of two-photon interference. Our results indicate that improvement is only possible in the absence of dephasing mechanisms, which typically set the quantum dot emission far from Fourier-transform limited. This result has an important implication: it points to the appropriate approach for designing photonic cavities that can improve the performance of quantum dot-based sources of entangled photon pairs. Namely, not only the biexciton emission rate must be modified to eliminate the cascade-induced correlations, but also the exciton emission rate must be modified to overcome the dephasing.

\begin{acknowledgments}
This work was supported by the CAPES/STINT project, grant No.
88887.646229/2021-01. J. L. was supported by the
Knut \& Alice Wallenberg Foundation (through the Wallenberg Centre for Quantum Technology (WACQT)). A.P. would like to acknowledge the Swedish Research Council (grant 2021-04494). C.S gratefully acknowledges funding from the German Ministry of Research and Education (BMBF) within the project EQUAISE. EQUAISE was funded within the QuantERA program.
\end{acknowledgments}

\end{document}